# BPMN Model to Smart Contract by Business Analyst

Christian Gang Liu
Faculty of Computer Science
*Dalhousie University*
Halifax, Canada
Chris.Liu@dal.ca

Peter Bodorik
Faculty of Computer Science
*Dalhousie University*
Halifax, Canada
Peter.Bodorik@dal.ca

Dawn Jutla
Sobey School of Business
*Saint Mary's University*
Halifax, Canada
Dawn.Jutla@gmail.com

*Abstract*— This paper addresses the challenge of creating smart contracts for applications represented using Business Process Management and Notation (BPMN) models. In our prior work we presented a methodology that automates the generation of smart contracts from BPMN models. This approach abstracts the BPMN flow control, making it independent of the underlying blockchain infrastructure, with only the BPMN task elements requiring coding. In subsequent research, we enhanced our approach by adding support for nested transactions and enabling a smart contract repair and/or upgrade. To empower Business Analysts (BAs) to generate smart contracts without relying on software developers, we tackled the challenge of generating smart contracts from BPMN models without assistance of a software developer. We exploit the Decision Model and Notation (DMN) standard to represent the decisions and the business logic of the BPMN task elements and amended our methodology for transformation of BPMN models into smart contracts to support also the generation script to represent the business logic represented by the DMN models. To support such transformation, we describe how the BA documents, using the BPMN elements, the flow of information along with the flow of execution. Thus, if the BA is successful in representing the blockchain application requirements using BPMN and DMN models, our methodology and the tool, called TABS, that we developed as a proof of concept, is used to generate the smart contracts directly from those models without developer assistance.

I. INTRODUCTION

The publication of the Bitcoin white paper in 2008 and the subsequent launch of the Bitcoin blockchain in 2009 sparked significant interest and research into blockchain technology. This emerging technology has garnered widespread attention from businesses, researchers, and the software industry due to its key attributes, such as trust, immutability, availability, and transparency. However, as with any new technology, blockchain and its associated smart contracts pose a range of challenges, particularly in areas like blockchain infrastructure and smart contract development.

Ongoing research is tackling several critical issues, including blockchain scalability, transaction throughput, and the high costs associated with consensus algorithms. In addition, smart contract development faces unique obstacles arising due to the blockchain infrastructure technology, such as a limited stack space, the oracle problem, data privacy concerns, support for long-running contracts, and cross-blockchain interoperability. These challenges have been the subject of extensive study, with numerous comprehensive literature reviews available [e.g., 1, 2].

The inherent constraints of blockchain technology complicate the development of smart contracts, as documented in several literature surveys [e.g., 3, 4]. Consequently, developers must not only be proficient in traditional software development but also possess expertise in smart contract programming for distributed environments, including the use of cryptographic techniques integral to blockchain infrastructure. To address these challenges and simplify smart contract development, research in [5-8], proposes leveraging Business Process Model and Notation (BPMN) models [9] as a foundation for generating smart contracts.

We also use BPMN to represent business application requirements, however, we take a different approach to transforming BPMN models into smart contracts. Our method leverages multi-modal modeling to represent the flow of business logic in a blockchain-agnostic manner, providing unique advantages for automated or semi-automated smart contract creation and deployment. As a proof of concept, we developed a tool called Transforming Automatically BPMN model into Smart contracts with Repair Upgrade (TABS+R), which automates the generation of smart contracts from BPMN models [10, 11].

It should be noted that the BPMN and DMN are standards created by the Object Management Group (OMG) [9]. Both are graphical standards that *have been designed to be readily understandable and used by both non-technical and technical people and thus form a bridge between the business and IT personnel*. BPMN is used to represent well-defined business processes, while DMN is used to specify business decisions and rules. The DMN standards specifies the use of the Friendly Enough Expression Language (FEEL) that was designed to write expressions in a way that is *easily understood by both business professionals and developers*. FEEL is used to define expressions in the context of BPMN and DMN diagrams [9].

As DMN and BPMN have been designed to be readily understood and used by business professionals, such as Business Analyst (BA), as well as IT personnel, we assume that a BA, who is responsible for requirements gathering for the blockchain application, is familiar with BPMN and DMN modeling. Consequently, as it is the BA who uses BPMN and DMN modeling to represent the blockchain distributed application, if we achieve automated transformation of BPMN models, for which DMN is used to express the business decision logic, we shall enable the BA to generate smart contracts without assistance by software developers as long as they can express the business logic using DMN.

*A. Objectives*

The main objective of this paper is to show the feasibility of generating methods of a smart contract from a BPMN model with business logic represented using DMN. To achieve the transformation, two separate subproblems must be addressed, namely (a) representation of the business logic in DMN and how it is transformed into the code executable by the generated smart contract, and (b) which information must be available for the transformation and how such information is represented in BPMN and DMN models.

*B. Contributions*

The main contributions of this paper include:
 i. Describing how the BA documents the flow of information along the flow of computation. This information is by the transformation of BPMN and DMN models into smart contracts.
 ii. Showing how the BPMN models are readily augmented with DMN modeling to represent the business logic of the blockchain application.
 iii. Proof of concept to show the feasibility of our approach to automated generation of smart contracts for applications modeled with BPMN and DMN.

*C. Outline*

In the second section, we outline our system architecture for creating smart contracts and for their execution and overview the significant features of our approach to automated development of smart contracts from BPMN and DMN models. The third section describes how the BA uses BPMN modeling to represent the flow of information to support the transformation, while the fourth section explains the use of DMN modeling to represent the business logic. The fifth section shows the tool in action on a selected use case. The last two sections provide related work and summary and conclusions, respectively.

## II. USING MULTI-MODAL MODELING FOR SMART CONTRACT GENERATION

In contrast to the other approaches to transforming BPMN models into smart contracts, we exploit multi-modal modeling to represent the flow of computation of the business logic in a blockchain-agnostic manner [10]. We subsequently extended our approach and methodology and created a PoC tool, called TABS+R [11, 12] to support:
- Semi-automated generation of smart contracts from BPMN models [10,11].
- Support for nested long running and multi-step transactions [11].
- Repair/upgrade of smart contracts [12].

The overall architecture of our system is illustrated in Fig. 1. It presents a block diagram outlining the key steps involved in transforming a BPMN model into smart contract methods. The diagram also includes a set of API methods (denoted as DAppAPI in Fig. 1) that facilitate interaction between a Distributed Application (DApp) and the smart contract methods. This architecture is typical of most approaches that generate smart contract methods from BPMN models [4-7]. In this setup, the DApp does not directly invoke the smart contract methods. Instead, it calls API methods provided by the *API-SCmethods* component in Fig. 1), which marshals the necessary parameters and then triggers the corresponding smart contract methods.

During the design phase, activities of actors involved in the smart contract are represented using multi-modal modeling. Concurrency is modeled using Discrete Event (DE) modeling, while functionality is represented with concurrent Finite State Machines (FSMs), forming a DE-FSM model. A key feature of this model is its blockchain-agnostic nature, meaning that the coordination of collaborative activities is described using DE modeling. Only the code for the BPMN task elements is blockchain-dependent, i.e., it needs to be written in a programming language that is supported by the target blockchain. For example, Ethereum-based blockchains typically use languages that produce code executed by the Ethereum Virtual Machine (EVM), whereas JavaScript or other languages may be used for scripting task elements in Hyperledger Fabric (HLF) blockchains.

Scripting the code for the BPMN task elements is relatively straightforward compared to scripting synchronization of collaboration of activities that is orchestrated through the transformation of BPMN models, with business logic represented using DMN modeling, into smart contracts. The code implementing a BPMN task is self-contained: In BPMN modeling, once the flow of computation enters a task element and its execution begins, the task completes its computation without interruption. Furthermore, the task code (i) can read information flowing into the task, (ii) read/write the blockchain state variables; and produce output information that flows out of the task when its computation finishes. Thus, the task code is self-contained in a smart contract method accessing only the state variables and the method's inputs and outputs, as represented by BA using the flow of information in a BPMN model as will be described in a following section. Consequently, the approach also leads to a modular design.

In summary, the flow of collaborative activities is modeled using DE-FSM modeling. The functionality of task elements is achieved by invoking methods that implement the business logic of each task. To coordinate these collaborative activities, a run-time monitor, implemented as a smart contract method deployed on the target blockchain, ensures the correct sequencing and execution of the activities. This monitor uses DE modeling to manage the invocation of individual activities, which are represented as methods within the monitor. Thus, if the target blockchain for the smart contract deployment has a monitor smart-contract deployed, the synchronization of the collaborative activities is blockchain agnostic. Furthermore, our approach deploys a monitor smart contract on the target blockchain automatically. In our proof of concept (PoC), the TABS+R tool, we implemented the monitor smart contract to be deployed on the Hyperledger Fabric (HLF) blockchain as well as on blockchains supporting the EVM [10-12].

III. BPMN MODELING BY BA

Before we describe BPMN modeling by the BA, we briefly overview information on storage of large files and communication between a smart contract and its external environment.

A. Preliminaries

As is the usual practice for blockchains, large document files or objects are not stored on the blockchain itself but are stored off-chain. For the storage of document files or large objects, we currently utilize the InterPlanetary File System (IPFS) [13] for its reliability and availability supported through replication.

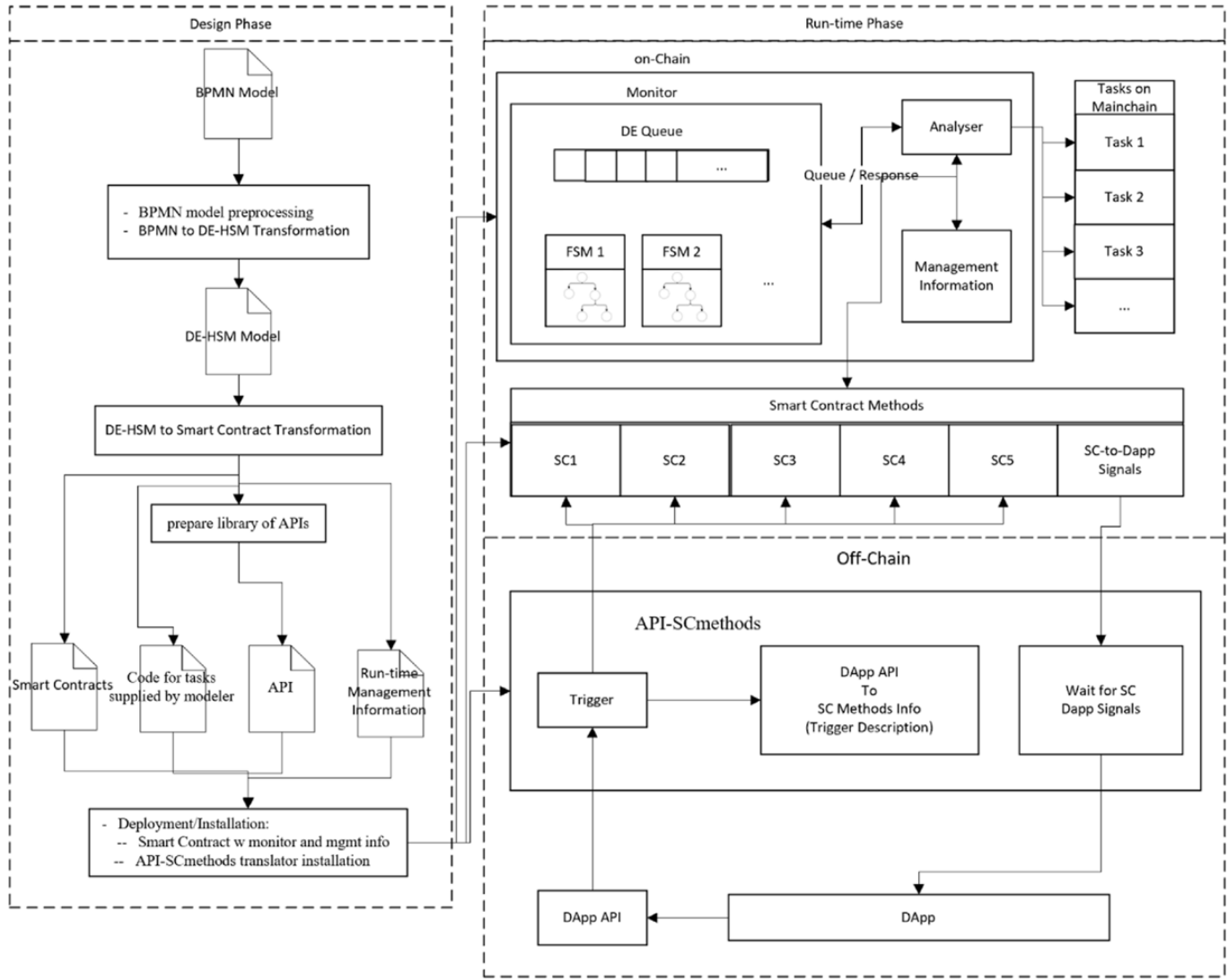

Fig. 1. System architecture for the design phase and for the execution phase (adopted from [10])

When a document is created and uploaded to IPFS, a new Content-addressed hash code IDentifier (CID) is generated. This CID is then signed and stored by the smart contract, providing a method to verify the document's authenticity, including confirming (i) authorship and (ii) immutability to ensure that the document has not been altered since its creation.

One of the key features that supports trust in smart contracts is that the methods within a smart contract do not have access to external resources, such as file systems or communication subsystems. Smart contract code can only access the state variables stored on the blockchain and the parameters passed to smart contract methods when they are invoked. Therefore, beyond the state variables, any additional information required by a smart contract must be marshaled by the *API-SCmethods* component before the smart contract method is invoked. The marshalled data is then passed as input parameters when invoking the smart contract methods.

Additionally, a smart contract must communicate the progress of its execution to the Distributed Application program (*DApp*) that invokes its methods. This is accomplished by emitting events from the smart contract methods, which are captured by the *API-SCmethods* component (as shown in Fig. 1) and relayed to the *DApp*.

*B. Exposition Use Case*

For explanatory purposes, we will use a simple BPMN model, shown in Fig. 2, that represents a sale of a large product, such as a combine harvester. The model shows that an agreement on the sale of the product is reached first, followed by arrangements for transporting the product. These transport arrangements include determining the requirements for transporting the product, such as safety measures for hazardous materials. Once the transport requirements are established, insurance and transport are arranged, and the product is shipped. After transportation, the product reception by the client is reviewed, and the payment is finalized.

Modeling is carried out by a Business Analyst (BA) who is assumed to be proficient in BPMN and DMN modeling, including the use of the FEEL language for decision logic. Additionally, we assume that the BA is familiar with JavaScript Object Notation (JSON), which is used to describe the flow of information throughout the computation process, as will be detailed later.

In Fig. 2, the first task, *RecAgr,* involves receiving a purchase offer document from an external source. Once accepted, the purchase agreement (i.e., a sales agreement) is passed to the next task, *GetTrReq*, for further processing. The sales agreement is represented by an associated data element, *SalesAgr*. The *GetTrReq* task determines the transport requirements for the product and stores them in a newly generated IPFS document, *TrRequirements.* This document is then passed to the subsequent processing step.

The transport requirements are forwarded to the *GetIns* and *GetTransp* tasks to secure insurance and a transporter, respectively. These tasks can be executed concurrently, as indicated by the fork gateway (diamond shape with a plus sign). The *GetIns* task generates the insurance contract, labeled *Insurance*, while the *GetTransp* task produces the *Transport* document.

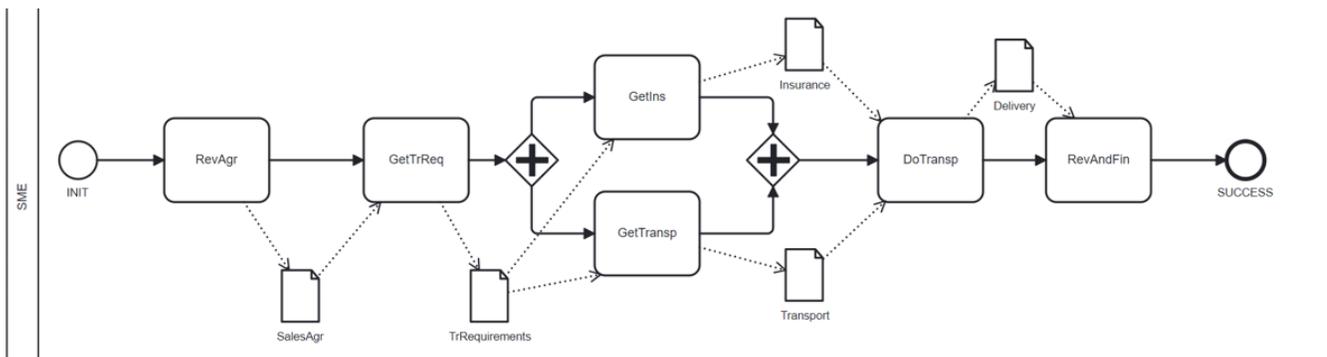

Fig. 2. BPMN model

Once both the insurance and transport contracts are obtained and provided to the transporter, the product is delivered to its destination, represented by the *DoTransp* task. Once the product is received by the purchaser, the reception of the product is recorded in the *Delivery* document that is forwarded to the final task, *RevAndFin*, that reviews and finalizes the contract.

Please note that the flow of activities shown in Fig. 2 is executed by a single actor, represented within one, in BPMN terminology, swimlane. This model is suitable for scenarios within organizations that lack sophisticated IT infrastructure, such as Small to Medium-sized Enterprises (SMEs). The simple use case is designed to demonstrate how the BA uses BPMN to document the flow of information along the computation process and how the BA applies DMN modeling to define the business logic.

In the following, we describe how a BA, working within the context of an SME, creates a BPMN model to track activities, document flows, and express the business logic decisions of BPMN task elements using DMN modeling.

*C. Documenting Flow of Information by BA*

The previous discussion, of the BPMN model in Fig. 2, illustrates the flow of computation, which is forked by a fork-gate, enabling the concurrent execution of the *GetIns* and *GetTransp* tasks. The figure also shows how the BA represents information as it flows along with the flow of computation. This is achieved by the BA documenting the transfer of information between tasks using an association object. In Fig. 2, the dotted arrows, from the *RecAgr* task to the *SalesAgr* association object and then from the *SalesAgr* to the *GetTrReq* task, indicate the transfer of the sales agreement information (*SalesAgr*) from the *RecAgr* task to the *GetTrReq* task.

We first describe how JSON is used to model the flow of information and then provide simple examples to clarify. To provide more details on the content of the *SalesAgr* document, the BA clicks on *the SalesAgr* icon to provide annotation about its contents.

Information flowing along the computation process flow may contain multiple items, each of which is described by an array of key-value pairs. For this purpose, the BA uses JSON to represent the information flowing along the computation process. Items, such as *item1* and *item2,* are represented as an array of JSON elements.

The first element in the array has the form: { "source": "string1" }. The value of "string1" can only be "file" or "http", denoting whether the information is sourced from a file or an HTTP service. If the value of string1, representing the value for the key "source", is "file", the next item in the array specifies the *CID* (Content Identifier) of the file from which the information is retrieved. This file is assumed to be in JSON format. The subsequent items in the array identify the fields (or components) within the file that need to be retrieved and passed as parameters to a smart contract method invoked by the *API-SCmethods* component.

If the value of string1 is "http", then there is an array of elements that contain information on (i) HTTP address of the service, (ii) input parameters, and (iii) output parameters containing the results of the service execution. The HTTP service is invoked with input parameters described, wherein the service returns information in its output parameters. Both the input and output parameters are described using the array elements. The HTTP service is invoked to implement the task and return the produced results in its output parameters that are recorded in the smart contract. For brevity, we will focus on describing how JSON is used to represent the content of files that provide information flowing along the computation process.

In Fig. 3, the file containing the relevant information is named *SalesAgr.json*, and its CID is provided. The array of elements within the JSON structure identify which components of the *SalesAgr.json* file are to be retrieved and passed as parameters to the smart contract method. In our simple use case, the JSON components to be retrieved and passed to the smart contract method include only the *product ID*, which is supplied to both *GetTrReq* and *GetIns* tasks. These tasks then use the product ID to retrieve further details about the product to be transported and then the requirements for its transport, if any.

This approach allows the BA to clearly define the flow of data in the smart contract system, ensuring smooth interaction between the BPMN and DMN models and the smart contract that is generated, and providing transparency and traceability in the overall process.

Information flowing into a task, as a result of invocation of a smart contract method, is prepared by the *API-SCmethods* component. It retrieves the information described by the JSON annotation of the *SalesAgr* association object, marshals it into the appropriate format, and passes it as input parameters to the smart contract method that implements the *GetTrReq* task.

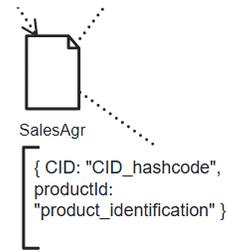

Fig. 3. Annotation to describe information flowing between tasks

For the subsequent sections, we assume that the monitor smart contract, required by the smart contracts generated by the TABS+R tool, has already been deployed on the target blockchain. We support currently either HLF blockchain or a blockchain that uses EVM.

## IV. DMN Modeling

We will use our simple example use case, represented by the BPMN model of Fig. 2. Assuming, for simplicity, that if the quoted price for the insurance is 15% or more of the product price, then the whole contract should be aborted due to the high cost. To make such a decision, the price of the product and the insurance cost need to be available.

To express the constraints on the insurance cost, we use the *business-rule task* element of BPMN. Functionally, the business-rule task first produces a value that is then forwarded to an exclusive gateway. The gateway uses the value, produced by the business rule task, to choose one of its forks for the outgoing flow of computation.

For our simple case, the business decision logic can be represented using a simple decision table as is shown in Fig. 4. Our tool invokes the graphical editor provided by Camunda (at Camunda.com) and available from BPMN.io. The decision table is created to check that the insurance quote, as a percentage of the price, is less than 15, in which case the next task to be executed is *DoTransp* to transport the product to its destination. The smart contract fails if the insurance quote, as a percentage of the price, is higher than 15%.

Once the decision table is completed, the business-rule task is represented by a rectangular icon with rounded corners that has a picture of a small table of rows and columns in the left top corner, as is shown in Fig. 5. From the business rule task there is an outgoing flow that contains the result of the business rule evaluation that is then used by the following exclusive fork gate to take one of the outgoing paths, one for the when the percentage is less than or equal to 15% that continues to the *DoTransp* task, while if the percentage is greater than 15%, the contract fails, resulting in automatic execution of recovery procedures as described in [11]

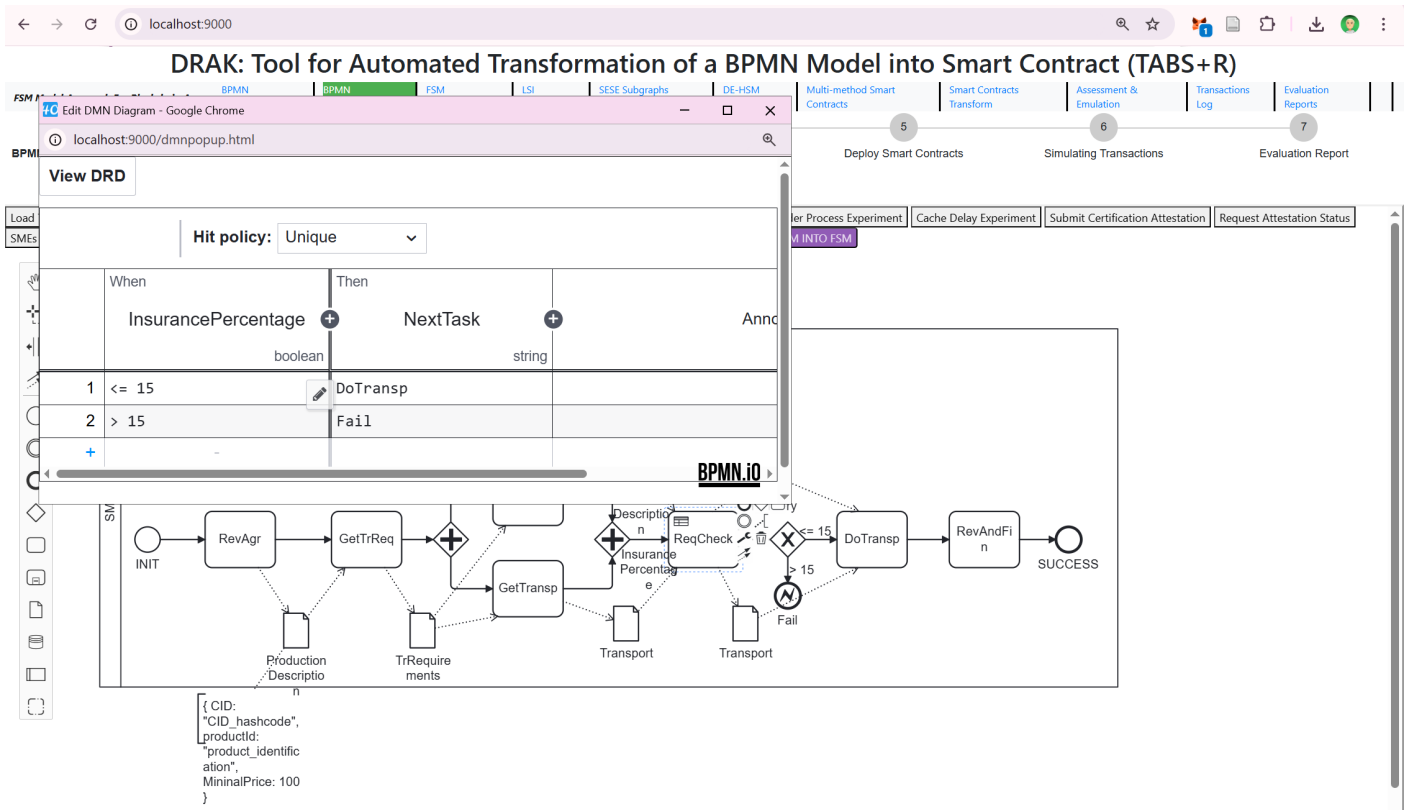

Fig. 4. Creating the decision table for a business-rule task

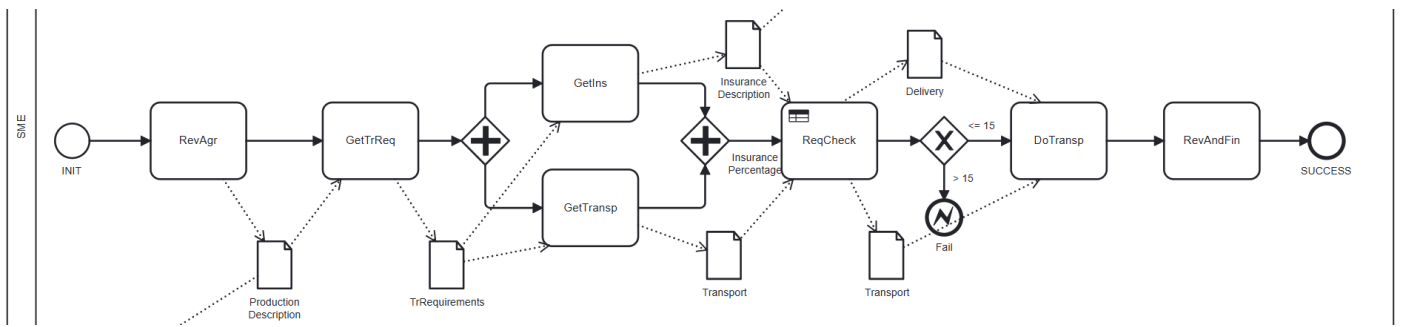

Fig. 5. BPMN model with the business-rule task

In addition to the simple decision tables, DMN modeling also incorporates the use of the Friedly Enough Expression Language (FEEL). FEEL was created by OMG as a part of DML using the following design principles with the aim to be a readable language for programmers and business analysts [ref to OMG doc or Camnuda tutorial]:

- Side-effect free
- Simple data model with numbers, dates, strings, lists, and contexts
- Simple syntax designed for a broad audience
- Three-valued logic (true, false, null)
- Control statements including assignment, conditional, looping, and range statements.
- Functions for string, numbers, data and time, and lists.

We acknowledge that currently we only support simple decision tables. However, FEEL has been implemented in BPMN modeling used for process orchestration, for instance by Camunda as described in BPMN.io, and we do not foresee design challenges.

## V. GENERATING SMART CONTRACTS BY BA IN SME

We analyzed a variety of use cases from the literature that focuses on transformation of BPMN models into smart contracts, such as use cases for Order-supply, Supply chains, Parts Order, Sales and Shipment, and Ordering Medications.

In each case, the creation, review, or amendment of these documents occurs off-chain. In such cases, the exchanged data between actors consists primarily of QR codes that identify the document files being shared, wherein the QR code is used as the documents unique ID that is analogous to the CID generated by the IPFS. The smart contract interactions among the partners are limited to the exchange of these documents, rather than directly handling the creation or modification of them.

Thus, when task executions can be performed off-chain, the task script code does not need to be provided on-chain, as long as the generation of the smart contracts from the BPMN model ensures a certified exchange of documents between on-chain and off-chain computations, which is readily supported by our approach as only CIDs are passed to the smart contract methods.

Operationally, in the absence of the IT support, the BA, or an operator trained by the BA, performs the actual activities represented by some of the tasks, while the smart contract records the result of the BA's activities. For instance, it is the BA who needs to execute the *GetTrReq* task. The BA needs the product description that is communicated by the BA to an insurance provider. The insurance provider communicates the insurance document to the BA who needs to store it in the file system to be accessible by the *API-SCmethods* component of the architecture shown in Fig. 1.

## VI. RELATED WORK

Several approaches to transforming BPMN models into smart contracts have been explored. The Lorikeet project focuses on transforming BPMN models into smart contracts to facilitate blockchain-based business process execution and asset management [5, 16]. The project employs a model-driven engineering approach, where BPMN models are analyzed and converted into smart contract methods that can be deployed on blockchain platforms, particularly Ethereum. An off-chain component is used to manage interactions between process participants and the blockchain, ensuring the execution of processes follows the predefined message exchanges in the BPMN model.

Additionally, Lorikeet supports asset control, enabling the management of both fungible and non-fungible assets, such as token registries and transfer methods, which are essential for business processes requiring asset handling. This approach allows for rapid prototyping, testing, and deployment of smart contracts based on BPMN models, enhancing flexibility and efficiency in blockchain-based business process automation [5, 16].

The Caterpillar project focuses on transforming Business Process Model and Notation (BPMN) models into smart contracts, providing a comprehensive architecture for executing business processes on the Ethereum blockchain [6, 7]. It adopts a three-layer architecture that includes a web portal, an off-chain runtime, and an on-chain runtime. The on-chain runtime layer is responsible for managing the execution of smart contracts that control workflow, interaction management, and process configurations based on the BPMN model. This approach ensures that business processes are executed transparently, securely, and efficiently within a blockchain environment.

The Caterpillar project emphasizes recording all business processes in a single pool, facilitating the management of interactions and ensuring the consistency of the process execution across multiple actors. By leveraging Ethereum as the blockchain platform, Caterpillar enables the seamless integration of BPMN models with decentralized applications, supporting the automation of business workflows through blockchain-based smart contracts [6, 7, 17].

The Collaborative Business Process Execution on Blockchain (CoBuP) project explores the transformation of BPMN models into smart contracts, offering a unique approach compared to traditional methods. CoBuP does not directly compile BPMN models into smart contracts [15]. Instead, it deploys a generic smart contract that invokes predefined functions based on the BPMN model, making it more flexible and adaptable to various process executions.

The CoBuP architecture is based on three layers: conceptual, data, and flow layers. BPMN models are first transformed into a JSON-based workflow model, which governs the execution of business processes by interacting with data structures on the blockchain. This allows for a decentralized, secure execution of business processes while maintaining the flexibility needed for collaborative environments. The project approach highlights the potential for blockchain to support complex business processes that require a high degree of collaboration, adaptability, and trust among participants.

## VII. SUMMARY AND CONCLUSIONS

In this paper, we first reviewed the progress on our project to transform the BPMN models into a smart contract. We then described our approach to augmenting the approach to allow modeling of business logic using DMN. We described how the BA annotates the BPMN model with information on the flow of data along the line of computation. This is required so that the business logic expressed using DMN can be mapped to the data/objects that are passed amongst the tasks of the BPMN model. We showed the use of DMN modeling to describe a simple example of expressing business logic using a decision table. Once the BA develops the BPMN and DMN models for the distributed applications, the BA uses our tool, TABS+R that we developed as a proof of concept, to transform the BPMN models, for which the simple business logic was expressed using DMN modeling, into a smart contract that is deployed on a target blockchain. Thus, we demonstrated that a BA uses the BPMN and DMN modeling to create models that are transformed into smart contracts without the assistance of a software developer. Of course, this is only for the case when the BA manages to express the business logic using DMN modeling.

It should be noted that DMN modeling is quite sophisticated as, in addition to the simple concept of the decision tables, it also includes the Friedly Enough Expression Language (FEEL) for representing the business rules/expressions with a simple data model and simple control constructs for conditionals, looping, and ranges. The language was designed to be understood by business professionals and IT personnel and thus should be friendly enough to be used by BAs.

Although the business-rule task is being proposed for BPMN and has already been used in software products, such as in modeling graphical editors presented in BPMN.io that we exploit in our proof of concept, it has not been yet officially approved. As a consequence, the current implementations of the models in engines powering the orchestration of the business processes may differ [18].

Although research progress is being made automated transformation of BPMN models into smart contracts, much work is needed before it can be applied to software product supporting the concept of Smart Contract as a Service (SCaaS), or more precisely (BPMN to Smart Contract) as a Service (*(BPMNtoSC)aaS*). Input to the service is the description of the BPMN and DMN models expressed in XML, and information on the target blockchain. Output from the transformation includes the methods of the smart contract deployed on the target blockchain. In addition, also output is the monitor smart contract, deployed on the target blockchain, that required for the coordination of the task activities.

As our TABS+R tool is only a proof of concept, we are focusing on verifying and validating the security of the smart contract methods generated by our approach. Although we use standard techniques to secure individual smart contract methods, the concept of a long-running transaction enforced by an automatically generated transaction mechanism [11] requires protection from the man-in-the-middle attacks.

In addition, for any transformation generating the smart contracts to be useful in production environment, appropriate plugins are required. We are currently augmenting our PoC tool to invoke HTTP services to automatically execute tasks, which would be useful for deployment of the smart contracts generated in organization supporting the use of HTTP services in their business processes.